\documentclass[a4paper]{jpconf}
\usepackage{graphicx}
\begin{document}
\title{CP violation and the H-A lineshape}

\author{ J.Bernab\'eu$^a$, D. Binosi$^{b}$, and J. Papavassiliou$^a$}

\address{$^a$ Departamento de F\`\i sica Te\`orica and IFIC
Centro Mixto, Universidad de Valencia-CSIC, E-46100, Burjassot,
Valencia, Spain }
\address{$^b$ ECT*, Villa Tambosi, Strada delle Tabarelle 286
I-38050 Villazzano (Trento), Italy}

\ead{joannis.papavassiliou@uv.es}

\begin{abstract}

In two-Higgs doublet models (and particularly in the MSSM) the CP-even
($H$) and CP-odd ($A$) neutral  scalars are nearly degenerate in mass,
and  their $s$-channel  production  would lead  to nearly  overlapping
resonances.  CP-violating effects may  connect these two Higgs bosons,
giving origin  to one-loop particle  mixing, which, due to  their mass
proximity,  can  be  resonantly  enhanced,  altering  their  lineshape
significantly.   We  show that,  in  general,  the  effect of  such  a
CP-violating mixing cannot  be mimicked by (or be  re-absorbed into) a
simple redefinition  of the  $H$ and  $A$ masses in  the context  of a
CP-conserving model.   Specifically, the effects of  the CP-mixing are
such that,  either the mass-splitting of  the $H$ and  $A$ bosons lies
outside the range  allowed by the theory in  the absence of CP-mixing,
and/or  the detailed energy  dependence of  the produced  lineshape is
clearly different from the one  obtained by redefining the masses, but
not allowing any mixing.

\end{abstract}

\section{Introduction}
In        the         
two-Higgs        doublet        models~\cite{Lee:1973iz}
and  in most  SUSY scenarios~\cite{Nilles:1983ge} the
extended  scalar sector  contains five  physical fields:  a  couple of
charged Higgs bosons  ($H^\pm$), a CP-odd scalar $A$,  and two CP-even
scalars  $h$ (the  lightest, which  is to  be identified  with  the SM
Higgs)  and $H$.  
It is expected that their discovery
and subsequenet study of their fundamental physical properties 
should be of central importance in the next decades.
In addition to the LHC, where their primary discovery might take place,
further detailed studies of their characteristics
have been proposed in recent years, most notably  
in the context of muon-\cite{Barger:1996jm} 
and photon colliders~\cite{Ginzburg:2001ph}.

A very characteristic feature of the the $H$-$A$ system is that 
the  masses $m_H$ and $m_A$ of these two particles are nearly degenerate.
Specifically, their tree-level mass eigenvalues are related by~\cite{Haber:1997dt}
\begin{equation}
m^2_{H} =\frac{1}{2}\left[M^2_Z+m^2_A + \sqrt{(M^2_Z+m^2_A)^2-4m^2_AM^2_Z\cos^22\beta}\right]\,.
\end{equation}
In the decoupling limit, $M_A\gg M_Z$, we have that
\begin{equation}
m^2_H  \approx  m^2_A + M_Z^2 \sin^2 2\beta,
\label{deg}
\end{equation}
which,  for $\tan\beta\ge2$  (and thus  $\cos^22\beta\approx1$), implies 
that $m_H\approx  m_A$.
In general, the inclussion of radiative corrections is known to modify 
significantly 
the above tree-level relation, mainly due to the 
large Yukawa coupling of the top-quark, 
but does {\it not} lift the mass degeneracy, 
especially in the parameter space
region where $m_A>2M_Z$ and  $\tan\beta\ge2$~ \cite{KKRW}.
Therefore, the 
high-resolution scanning  of  the lineshape  of  the $H$-$A$ system is
expected   to  reveal   two   relatively  closely   spaced,  or   even
superimposed, resonances~\cite{Barger:1996jm}.

Evidently, if the CP  symmetry is exact, the $H$  cannot mix with the $A$,  at any given
order.  However,  in  the  presence  of  a  CP-violating  interactions
\cite{Bernabeu:1987gr,Pilaftsis:1998dd}
the $H$  can mix  with  the $A$
already  at one-loop  level,  giving rise  to  a non-vanishing  mixing
self-energy  $\Pi_{HA}(s)$.  
Such mixing,  in  turn,  can be  measured
through  the   study  of  appropriate  CP-odd   observables,  for example 
left-right     asymmetries.    
As     has     been    explained     in
\cite{Pilaftsis:1997dr} the CP-violating  amplitude  is
particularly enhanced  near resonace, if the two  mixing particles are
nearly degenerate,  a condition which  is naturally fullfilled  in the
$H$-$A$ system.   
Furthermore, the mixing  between $H$ and $A$  has an
additional profound effect for the  two masses: the near degeneracy of
the two  particles is lifted, and  the pole masses  move further apart
\cite{Carena:2001fw};   as  a   result,  the   originally  overlapping
resonances of the CP-invariant theory tend to be separated.

In  \cite{Bernabeu:2006zs}   we  have  explored   the  possibility  of
detecting the  presence of CP-mixing  between $H$ and $A$  through the
detailed  study  of  the  cross-section  of  the  $s$-channel  process
$\mu^+\mu^-\to A^*,H^*\to  f\bar f$ as function  of the center-of-mass
energy.   Although the  lineshape  is a  CP-even  quantity, there  are
certain  characteristics that  signal the  presence of  a CP-violating
mixing.  Due to the facts reviewed above,  when  studying  the
lineshape of the $H$-$A$ system, one may envisage two, physically very
different, scenarios.   In the  first one, the  CP symmetry  is exact,
with the position of  resonances determined by Eq.(\ref{deg}) plus its
radiative corrections; the relative position between the two resonance
will then specify $\tan\beta$. In the second scenario, CP is violated,
resulting  in mixing  at  one-loop level  between  CP-even and  CP-odd
states  wich  translates  into  a non-vanishing  off-shell  transition
amplitude $\Pi_{HA}(s)$. Then, for  the same mass splitting $\delta m$
as in the previous case, one may not reach the same conclusion for the
value of $\tan\beta$, because one  could have started out with the two
masses  almost  degenerate, corresponding  to  a  different value  for
$\tan\beta$  (lower or  higher depending  on  the value  of the  $\mu$
parameter), and  the observed  separation between $m_{H}$  and $m_{A}$
may be due  to the lifting of the degeneracy  produced by the presence
of the aforementioned $\Pi_{HA}(s)$.

\begin{figure}[h]
\includegraphics[scale=0.97]{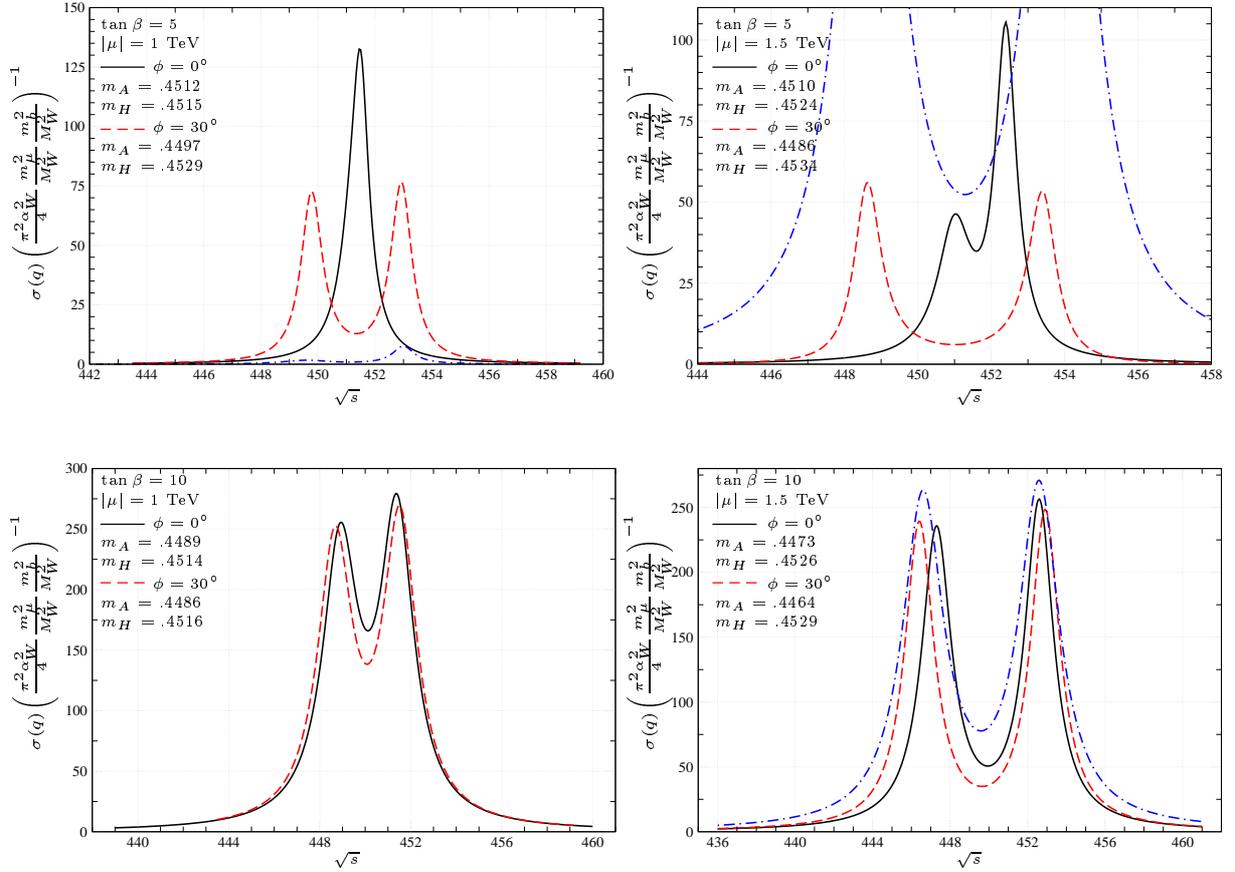}
\caption{\small{\it The $H$-$A$ lineshape for different values of $|\mu|$,
$\tan\beta$ and $\phi$. The  black continuous curves correspond to the
CP-invariant  limit  of the  theory  ($\phi=0^\circ$),  the  red
dashed to  the case where CP-breaking  phases have
been  switched on (with  $\phi=30^\circ$). Finally,  the dashed-dotted
blue curve, when present, corresponds  to the CP-invariant limit of the
theory  re-calculated   for  a  different  value   of  $\tan\beta$  to
accommodate the mass-splitting observed in the CP-breaking case. }}
\end{figure}

\section{Results}

We have studied the lineshape of the process 
$\mu^+\mu^-\to A^*,H^*\to  f\bar f$, assuming the presence 
of CP-violating one-loop mixing  between $H$ and $A$, 
induced by the Yukawa couplings to the top and 
bottom squarks~\cite{Pilaftsis:1998dd}.
The kinematic regime considerd is $m_A>2M_Z$
and   $\tan\beta\geq2$; 
the rotation  angle of the scalar  top and bottom quarks  are equal to
$\pi/4$,  and we
assume          universal         squarks          soft         masses
($\tilde{M_Q}=\tilde{M_t}=\tilde{M_b}=\  M_0=0.5$  TeV)  and  trilinear
couplings  ($A_t=A_b=A$ with  $|A|=1$).  Finally  $m_{H^\pm}$  has been 
fixed at the value of $0.4571$ TeV.

The result  of our  analysis may be summarized 
as follows~\cite{Bernabeu:2006zs}:
In the cases when the CP-breaking
phases are sizeable  ($\phi=90^\circ$, not shown), one
can always clearly distinguish between the CP-invariant and CP-beaking
scenarios, either  because the  mass splitting in  the latter  case is
just too big for being due to CP-invariant radiative corrections only,
or  because   of  the  different   energy  dependence  of   the  cross
section.   For  smaller   values  of   the  CP-breaking   phases  (say
$\phi=30^\circ$, see Fig.1), 
one  can still distinguish  between the CP-invariant
and   CP-breaking  case   only   when  $\tan\beta$   is  small;   when
$\tan\beta>10$  one cannot tell the two cases apart simply by 
studying the lineshape. Thus, in general we conclude that
either the mass-splitting of the  H and A
Higgs boson, in a CP-mixing  scenario, cannot be accounted for 
in  absence  of  CP-mixing, and/or  the  detailed  energy
dependence  of  the line-shape  allows  to  discriminate between  both
scenarios.

We therefore believe that the experimental determination of the $H$-$A$
line-shape, in conjunction with  CP-odd asymmetries and other suitable
observables, may  provide valuable information for  settling the issue
regarding CP-mixing effects in two-Higgs doublet models.
It would be interesting to extend this analysis to the 
more complicated case of the LHC, 
taking into account the
variable gluon virtualities, in the spirit of~\cite{Godbole:2007cn}.

\subsection{Acknowledgments}
This work was supported by the Spanish MEC under the grant FPA 2005-01678
and the Fundaci\'on General of the University of Valencia.

\end{document}